\documentclass[12pt,a4paper,twoside]{article}
\usepackage{epsfig}

\long\def\ifundef#1#2#3{\expandafter\ifx\csname
  #1\endcsname\relax#2\else#3\fi}
\ifundef{ensuremath}{\newcommand{\ensuremath}[1] {\relax\ifmmode {#1}\else {$#1$} \fi}}{}
\ifundef{mathcal}{\newcommand{\mathcal}[1]{{\cal #1}}}{}
\ifundef{mathrm}{\newcommand{\mathrm}[1]{\rm #1}}{}
\ifundef{resizebox}{\newcommand{\resizebox}[3] 
{#3 
}}{}
\ifundef{includegraphics}{\newcommand{\includegraphics}[1] 
{$#1$ 
\typeout{PostScript File #1 not included!}
}}{}

\newcommand{\epem}   {\ensuremath{\mathrm{e^+e^-}}}

\newcommand{\as}     {\ensuremath{\alpha_s}}

\newcommand{\asq}    {\ensuremath{\alpha_s(Q)}}
\newcommand{\asmz}   {\ensuremath{\alpha_s(M_{\mathrm{Z^0}})}}

\newcommand{\oaa}    {\ensuremath{\mathcal{O}(\alpha_s^2)}}

\newcommand{\znull}  {\ensuremath{\mathrm{Z^0}}}

\newcommand{\mz}     {\ensuremath{M_{\mathrm{Z^0}}}}

\newcommand{\chisq}  {\ensuremath{\chi^2}}
\newcommand{\chisqd} {\ensuremath{\chi^2/\mathrm{d.o.f.}}}
\newcommand{\xmu}    {\ensuremath{x_{\mu}}}

\ifundef{pt}{ }
            { }

\newcounter{hours}
\newcounter{minutes}

\newcommand{\Printtime}{%
  \setcounter{hours}{\time/60}%
  \setcounter{minutes}{\time-\value{hours}*60}%
  \ifthenelse{\value{hours}<10}{0}{}\thehours:%
  \ifthenelse{\value{minutes}<10}{0}{}\theminutes}

\begin{document}
%
\begin{titlepage}
\vspace*{-10mm}
\hbox to \textwidth{ \hsize=\textwidth
\hspace*{0pt\hfill} 
\vbox{ \hsize=58mm
{
\hbox{ PITHA 98/21\hss}
\hbox{ July 02, 1998\hss } 
}
}
}

\bigskip\bigskip\bigskip
\begin{center}
{\Huge\bf
Measurement of {\ensuremath{C}}-Parameter \\[1.5mm]
and Determination of {\ensuremath{\alpha_s}} from \\[1.5mm]
{\ensuremath{C}}-Parameter and Jet Broadening \\[1.5mm]
at PETRA Energies
}

\end{center}
\bigskip\bigskip
\begin{center}
P.A.~Movilla~Fern\'andez$^{(1)}$, O.~Biebel$^{(1)}$, S.~Bethke$^{(1)}$ \\
 and the JADE Collaboration$^{(2)}$
\end{center}
\bigskip

\begin{abstract}
\noindent 
 e$^+$e$^-$ annihilation data recorded by the JADE detector
 at PETRA were used to measure the $C$-parameter for the
 first time at $\sqrt{s}= 35$ and $44$ GeV. The distributions
 were compared to a resummed QCD calculation which
 recently became available for this observable.
 In addition, we applied extended resummed calculations to 
 the heavy and wide jet broadening variables, $B_T$ and $B_W$,
 which now include a proper treatment of the quark recoil 
 against multi-gluon emission with single-logarithmic accuracy.
 We further investigated power corrections to the mean 
 values of the observables mentioned above. In this study, 
 we considered all available e$^+$e$^-$ data between 
 $\sqrt{s}= 35 $ and $172$ GeV.
\end{abstract}

\vspace*{0pt\vfill}
\vfill
\bigskip\bigskip\bigskip\bigskip

{
\small
\noindent
$^{(1)}$ 
\begin{minipage}[t]{155mm} 
III. Physikalisches Institut der RWTH Aachen,
D-52056 Aachen, Germany \\
contact e-mail: Otmar.Biebel@Physik.RWTH-Aachen.DE
\end{minipage}\\
$^{(2)}$ 
\begin{minipage}[t]{155mm} 
for a full list of members of the JADE Collaboration see Reference \cite{bib-naroska}
\end{minipage}
\hspace*{0pt\hfill}
}

\end{titlepage}
%
%
\newpage
\section{ Introduction }
In a recent publication~\cite{bib-newJADE} we have presented a study of 
event shapes observables and determinations of \as\ using data of 
\epem\ annihilations at $\sqrt{s} = 22$ to $44$~GeV recorded with the
former JADE detector \cite{bib-JADEdet} at the PETRA collider. 
This study provided valuable information which was not available before 
from $\epem$-annihilations in the PETRA energy range. The results on \as,
obtained in a similar manner as those from the experiments at LEP, 
demonstrated that the energy dependence of \asq\ is in good agreement 
with the prediction of Quantum Chromodynamics (QCD). Evolved to the 
$\znull$ mass scale, the results are in good agreement with those 
obtained at LEP, and are of similar precision. In addition, power 
corrections, applied to analytic QCD calculations of the mean
values of event shape distributions, were found to qualitatively
and quantitatively match the effects of hadronisation. Thus QCD could
be tested without the need of phenomenological hadronisation models.

Meanwhile the perturbative calculations for the jet broadening
variables were improved by including a proper treatment of quark 
recoil~\cite{bib-new-jetbroadening}. Furthermore for the $C$-parameter
a resummation of leading and next-to-leading logarithm terms to all orders
of \as\ (NLLA) became available~\cite{bib-C-resummation}. Beside these
advances in the perturbative description of event shape observables
progress was made in the understanding of non-perturbative power
corrections to the event shape observables and their mean values.
In particular two-loop calculations of such corrections were
performed \cite{bib-Milan-factor} which modify the one-loop result 
of the power correction to the event shapes by a factor (Milan factor).

In this paper we complement our previous publication by
a new \oaa+NLLA determination of \as\ from $C$-parameter and update 
the determination from jet broadening at $\sqrt{s} = 35$ and $44$ GeV
in the Sections~\ref{sec-procedure} and \ref{sec-alphas}.
We also applied power corrections to the $C$-parameter distributions
and re-investigated those for the mean values of the thrust, heavy
jet mass and both jet broadening 
observables in Section~\ref{sec-meanvalues}. We start in Section~\ref{sec-data}
with a brief summary of the data samples used and draw conclusions from our 
results in Section~\ref{sec-conclusions}.

\section{ Data samples and Monte Carlo simulation }
\label{sec-data}
We analysed 
data recorded with the JADE detector in 1984 to 1986 at 
centre-of-mass 
energies of $39.5$-$46.7$~GeV and around $35$~GeV.
The JADE detector was operated from 1979 until 1986 at the 
PETRA electron-positron collider at centre-of-mass energies of
$\sqrt{s} = 12$ to $46.7$ GeV. 

A detailed description of the JADE detector can be found
in~\cite{bib-naroska,bib-JADEdet}. 
The main components of the detector were the
central jet chamber to measure charged particle tracks
and the lead glass calorimeter to measure energy depositions of
electromagnetic showers, which both
covered almost the whole
solid angle of $4\pi$. 

Multihadronic events were selected by the standard JADE selection 
cuts~\cite{bib-JADEtrigger}. 
All charged particle tracks, assumed to be pions, with a total 
momentum of $|\vec{p}| > 100$~MeV$/c$ were considered in the 
analysis. Energy clusters in the
electromagnetic calorimeter, assumed to be photons, were considered 
if their energies exceeded $150$ MeV after correction for 
energy deposited by associated tracks.  

Background from two-photon processes and 
$\tau$-pair events and from events with 
hard initial state photon radiation
were removed by cuts on the visible energy 
$E_{\mathrm{vis}} = \sum E_i$, 
the total missing momentum
$p_{\mathrm{miss}} = |\sum \vec{p}_i|$ ($\vec{p}_i$ and $E_i$
are the 3-momentum and the energy of the tracks and clusters), 
the longitudinal balance relative to the \epem\ beam axis of momenta 
$p_{\mathrm{bal}} = |\sum p^z_i/E_{\mathrm{vis}}|$
and the polar angle of the
thrust axis, $\theta_{T}$:
\begin{itemize}
\item $E_{\mathrm{vis}} > \sqrt{s}/2$~;
\item $p_{\mathrm{miss}} < 0.3\cdot \sqrt{s}$~;
\item $p_{\mathrm{bal}} < 0.4$~;
\item $|\cos\theta_{T}| < 0.8$~.
\end{itemize}
Thus the backgrounds from $\gamma\gamma$ and $\tau$-pair events were reduced 
to less than $0.1\%$ and $1\%$, respectively \cite{bib-JADEeventsel}.
The numbers of events which were retained after these cuts for this 
analysis are listed in Table~\ref{tab-eventnumbers}.

\begin{table}
\begin{center}
\begin{tabular}{|c|c||c|c|}
\hline
year & $\sqrt{s}\ [\mathrm{GeV}]$ & data & MC \\ 
\hline\hline
1984/85 &  $40$-$48$ &             $\ 6158$ & $14\thinspace 497$  \\ \hline
   1986 &  $35$      &   $20\thinspace 926$ & $25\thinspace 123$  \\
\hline
\end{tabular}
\end{center}
\caption{\label{tab-eventnumbers}
Number of events in data and in Monte Carlo detector simulation retained after 
application of the multihadron selection cuts described in the text. 
}
\end{table}

Corresponding original Monte Carlo detector simulation data 
for $35$ and $44$~GeV
were based on the QCD parton shower event generator
JETSET 6.3~\cite{bib-JETSET}.
These original Monte Carlo events at 
$35$~GeV had the coherent branching for the parton 
shower while the $44$~GeV events had non-coherent branching
\footnote{The different treatment of coherence in these samples of
simulated data has no visible influence on the results of this study.
}.
The main parameters used for event
generation are given in Section~\ref{subsec-systematics}.
Both samples included a simulation of the acceptance and resolution
of the JADE detector.

\section{Experimental procedure}
\label{sec-procedure}

From the data samples described in the previous section, the
event shape distribution of the $C$-parameter was determined.
The $C$-parameter is defined as~\cite{bib-C-parameter}
\begin{displaymath}
            C = 3 (\lambda_1 \lambda_2  + \lambda_2 \lambda_3  + \lambda_3 \lambda_1 )
\end{displaymath}
where $\lambda_\gamma$, $\gamma=1, 2, 3$, are the eigenvalues of the momentum tensor
\begin{displaymath}
     \Theta^{\alpha\beta} = \frac{\sum_i \vec{p}_i^{\,\alpha} \vec{p}_i^{\,\beta} / |\vec{p}_i|}
                                 {\sum_j |\vec{p}_j|}
   \ \ \ .
\end{displaymath}

\subsection{Correction procedure}
\label{subsec-correction}
Limits of the detector's acceptance and resolution and effects due
to initial state photon radiation were corrected by applying a 
bin-by-bin correction procedure to the event shape distributions.
The correction factors were 
defined by the ratio of the distribution calculated from events
generated by JETSET 6.3
at {\em hadron level} over the same distribution at {\em detector
level}.
The {\em hadron level} distributions were obtained from JETSET 6.3
generator runs without detector simulation and without initial state
radiation, using all particles with lifetimes $\tau > 3\cdot
10^{-10}$~s.
Events at {\em detector level} contained initial state photon
radiation and a detailed simulation of the detector response, and 
were processed in the same way as the data.

Next, the data distributions were further corrected for hadronisation 
effects. 
This was done by applying bin-by-bin correction factors derived from
the ratio of the distribution at {\em parton level} over the same
distribution at {\em hadron level}, which were 
calculated from JETSET generated
events before and after hadronisation, respectively. 
The correction factors are typically of the order $10$ to $20$\%
growing large towards the $2$-jet region. In the case of the
$C$-parameter the correction factors are also large next to the 
$3$-jet boundary which is at $C=0.75$.
The data distributions, thus corrected to the {\em parton level}, 
can be compared to analytic QCD calculations.

\subsection{Systematic uncertainties}
\label{subsec-systematics}
Systematic uncertainties of the corrected data distributions were
investigated by modifying details of the event selection and of the 
correction procedure. For each variation the whole analysis was repeated 
and any deviation from the main result was considered a systematic
error. 
In general, the maximum deviation from the main result for each 
kind of variation was regarded as symmetric systematic 
uncertainty.
The main result was obtained using the default selection and 
correction procedure as described above.

In detail, we considered either tracks or clusters only for the measurement 
of the event shape distributions. We varied the cut
on $\cos\theta_T$ by $\pm 0.1$. The cut on $p_{\mathrm{miss}}$ was
either removed or tightened to $p_{\mathrm{miss}} < 0.25 \cdot \sqrt{s}$. 
Similarly, the momentum balance requirement was either restricted to 
$p_{\mathrm{bal}} < 0.3$ or dropped. We also varied the cut
for the visible energy $E_{\mathrm{vis}}$ by $\pm 0.05 \cdot \sqrt{s}$.
In order to check the residual contributions from $\tau$-pair events we
also required at least seven well-measured charged tracks.

To study the impact of the hadronisation model of the JETSET 6.3
generator, the values of several significant model parameters were
varied around their tuned values from Reference~\cite{bib-JADEtune}
used for our main result. 
%
Different sets of correction factors to correct the data from {\em
hadron level} to {\em parton level} were generated by varying single
parameters of the JETSET generator.
The variations were chosen to be similar to the one 
standard deviation percentage limits
obtained by the OPAL Collaboration from a parameter 
tuning of JETSET at $\sqrt{s} = \mz$~\cite{bib-OPALtune}. 
In detail, we investigated the effects due to 
parton shower, hadronisation parameters, and quark masses. 
The amount of gluon radiation during the parton shower development was modified by varying 
$\Lambda_{\mathrm{LLA}}$ by $\pm 50$~MeV around the tuned value $400$~MeV. 
To vary the onset of hadronisation, we altered the parton shower cut-off 
parameter $Q_0$ by $\pm 0.5$~GeV around the tuned value of $1$~GeV. We 
used the full observed variation of 
\as\ to reflect a variation of $Q_0$ between $0$ and $2$~GeV.
The width $\sigma_0=300$~MeV of the transverse momentum distribution 
in the
hadronisation process was varied by $\pm 30$~MeV.
The LUND symmetric fragmentation function, applied to hadronise
events of up, down and strange quarks,  was varied by changing the 
$a$ parameter from $0.5$ by $\pm 0.225$ whereas the $b$ parameter was 
kept fixed at $0.9$. As a systematic variation we used the LUND~\cite{bib-JETSET}
instead of the Peterson et al.~\cite{bib-Peterson} fragmentation function 
for charm and bottom quarks.
The effects due to the bottom quark mass were studied by 
restricting the model calculations which were used to determine the
correction factors
to up, down, 
strange, and charm quarks (udsc) only. 
Any deviation from our main result due to mass effects
was treated as asymmetric error.  

\section{ Determination of \as }
\label{sec-alphas}
\subsection{ Corrected event shape distributions}
After applying the corrections for detector and for initial state radiation 
effects we obtained the $C$-parameter event shape distributions at {\em hadron 
level}. 
In Tables~\ref{tab-eventshapes-35+44GeV}
the corrected data values are listed with
statistical errors and experimental systematic uncertainties. The 
mean values of the distributions are also given.
Our measured results for the
jet broadening event shape distributions can be found in 
Ref.~\cite{bib-newJADE}.

  
\subsection{ Determination of \as\ using \oaa+NLLA calculations}
We determined \as\ by \chisq\ fits to event 
shape distributions of $C$ and also $B_T$ and $B_W$
corrected to the {\em parton level}. 
For the sake of direct comparison to other published results 
we chose the so-called ln($R$)-matching
scheme to merge the \oaa\ with the NLLA calculations. The fits to the
$C$-parameter distributions applied the resummation results obtained 
in~\cite{bib-C-resummation}. For the fits to the jet broadening measures 
we used the improved calculation of Ref.~\cite{bib-new-jetbroadening} which 
includes a proper treatment of the quark recoil against an ensemble of soft 
gluons that is essential in the calculation of the jet broadening distributions.
The renormalisation scale factor, $\xmu \equiv \mu/\sqrt{s}$, was
set to $\xmu = 1$ for
the main result. 
Here, the value of $\mu$ defines 
the energy scale at which the theory is renormalised. 

\begin{figure}[!htb]
\vspace*{-7mm}
\begin{center}
\resizebox{79mm}{!}{\includegraphics{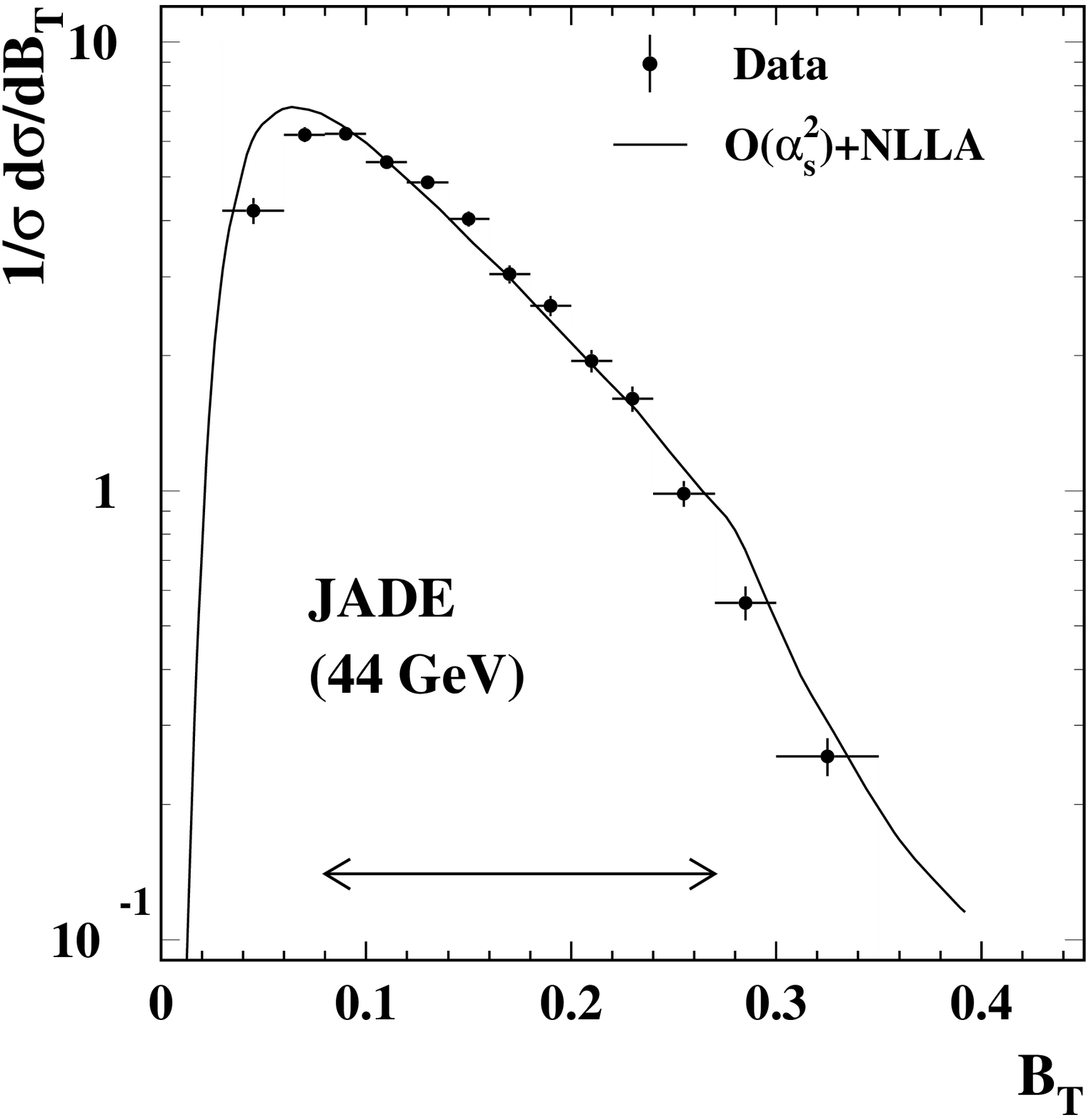}}
\resizebox{79mm}{!}{\includegraphics{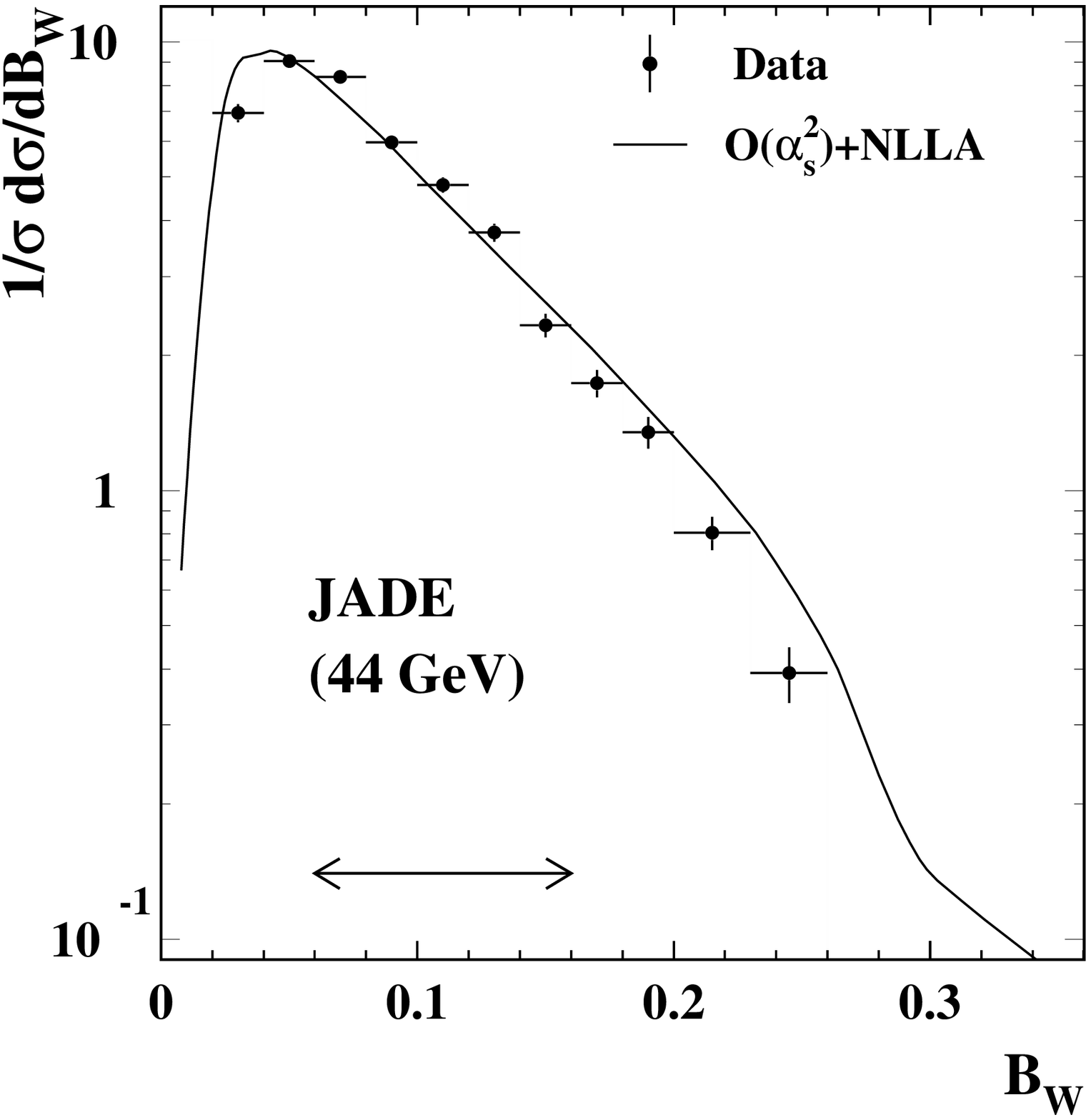}}\\
\resizebox{79mm}{!}{\includegraphics{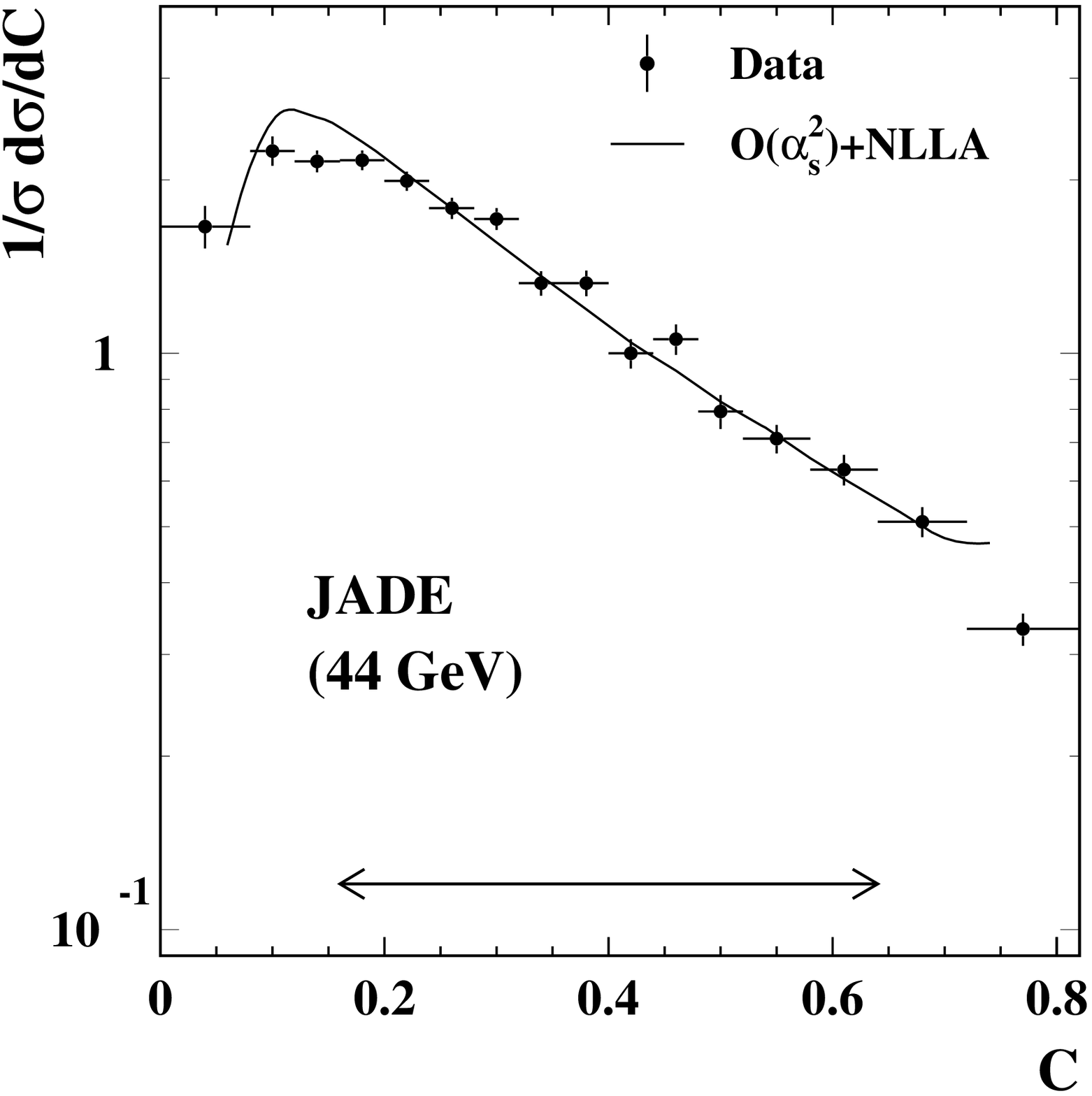}}
\end{center}
\caption{\label{fig-asresult-44GeV}
Distributions measured at $\protect\sqrt{s} = 44$~GeV and corrected 
to parton level are shown for the total and wide jet broadening $B_T$ and $B_W$
and for the $C$-parameter.
The fits, using the improved \oaa+NLLA QCD prediction for the jet broadening 
are overlayed and the fit 
ranges are indicated by the arrows. 
The error bars represent statistical errors only.
}
\end{figure}

\begin{figure}[!htb]
\vspace*{-7mm}
\begin{center}
\resizebox{79mm}{!}{\includegraphics{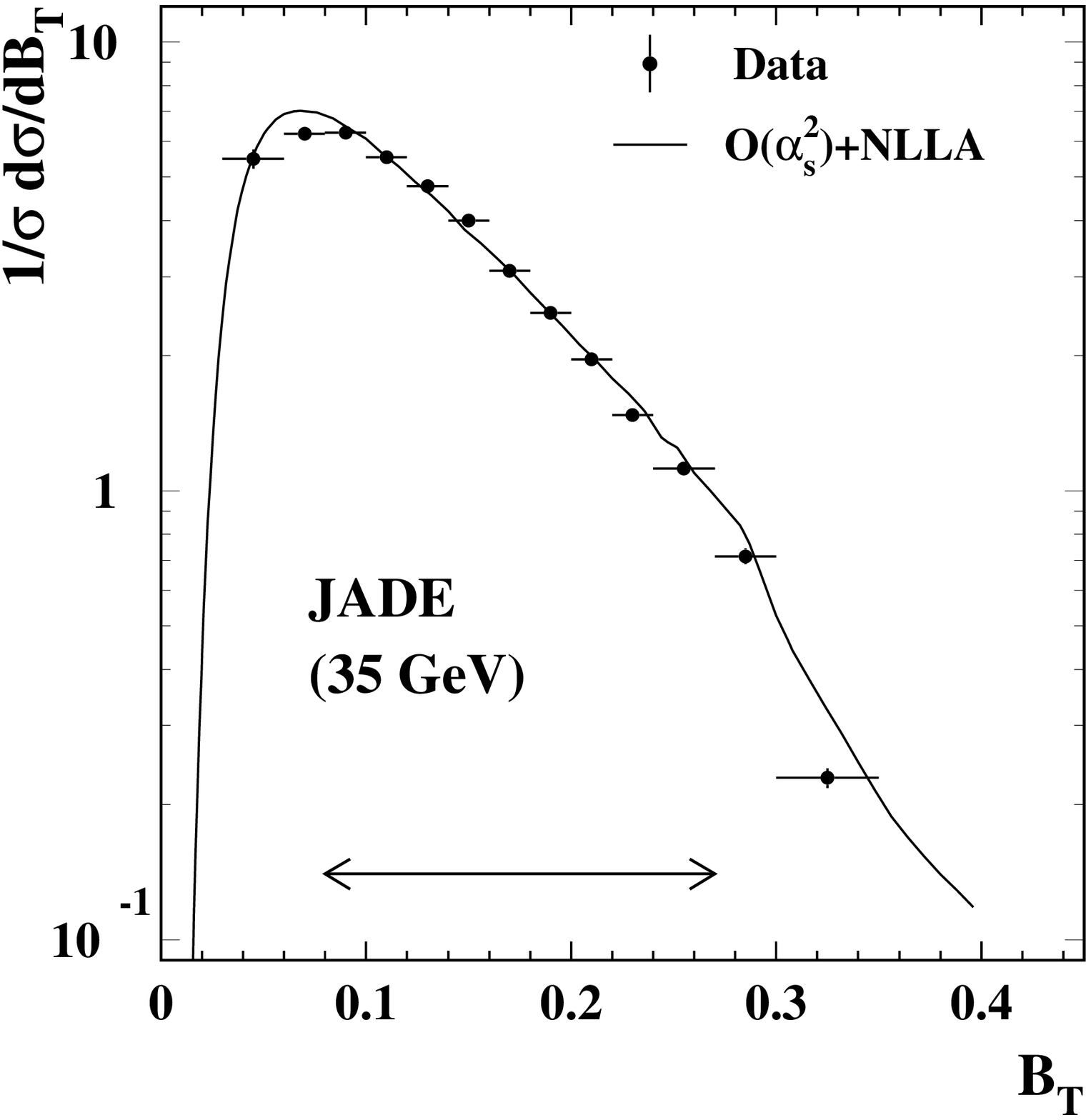}}
\resizebox{79mm}{!}{\includegraphics{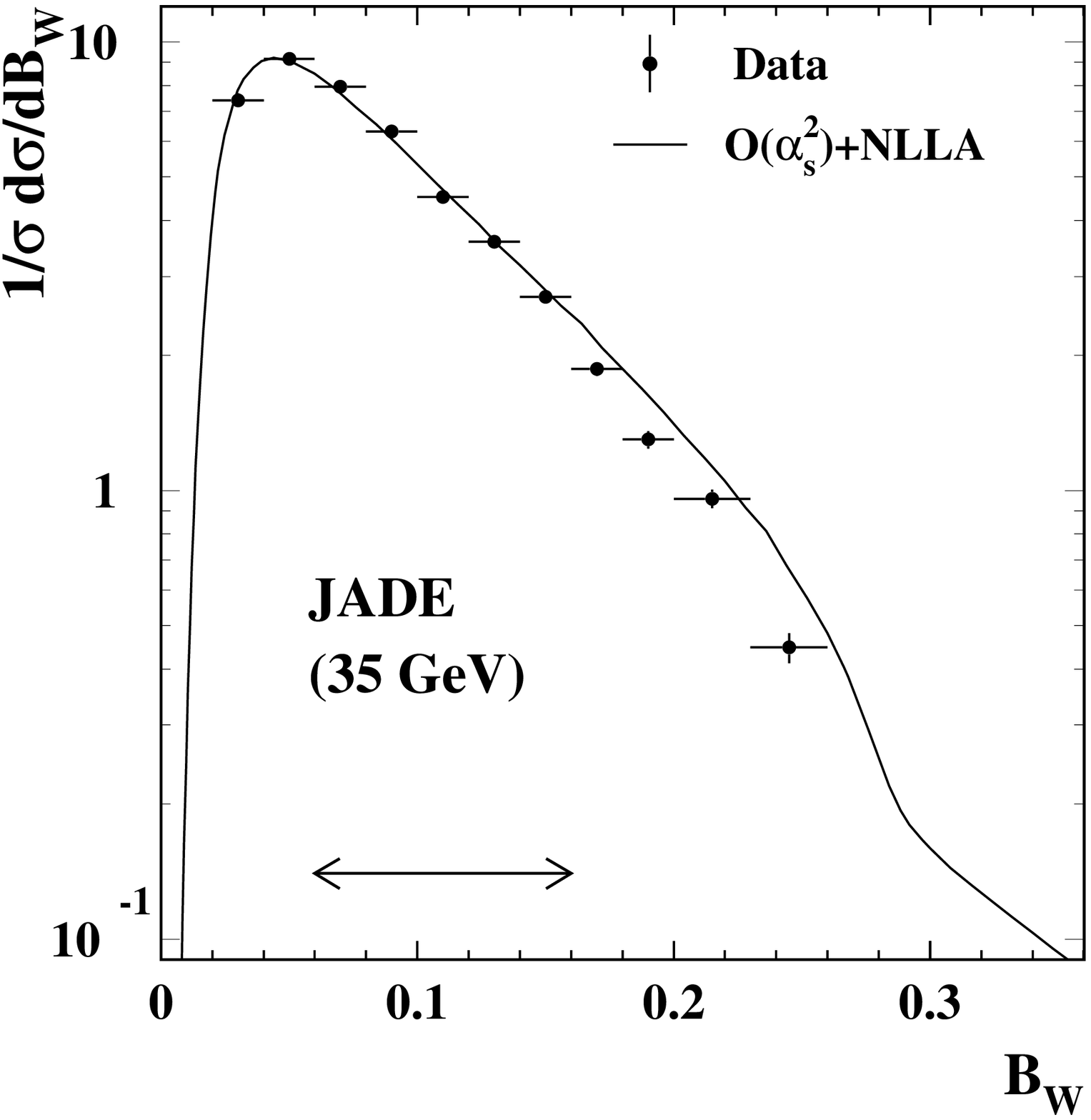}}\\
\resizebox{79mm}{!}{\includegraphics{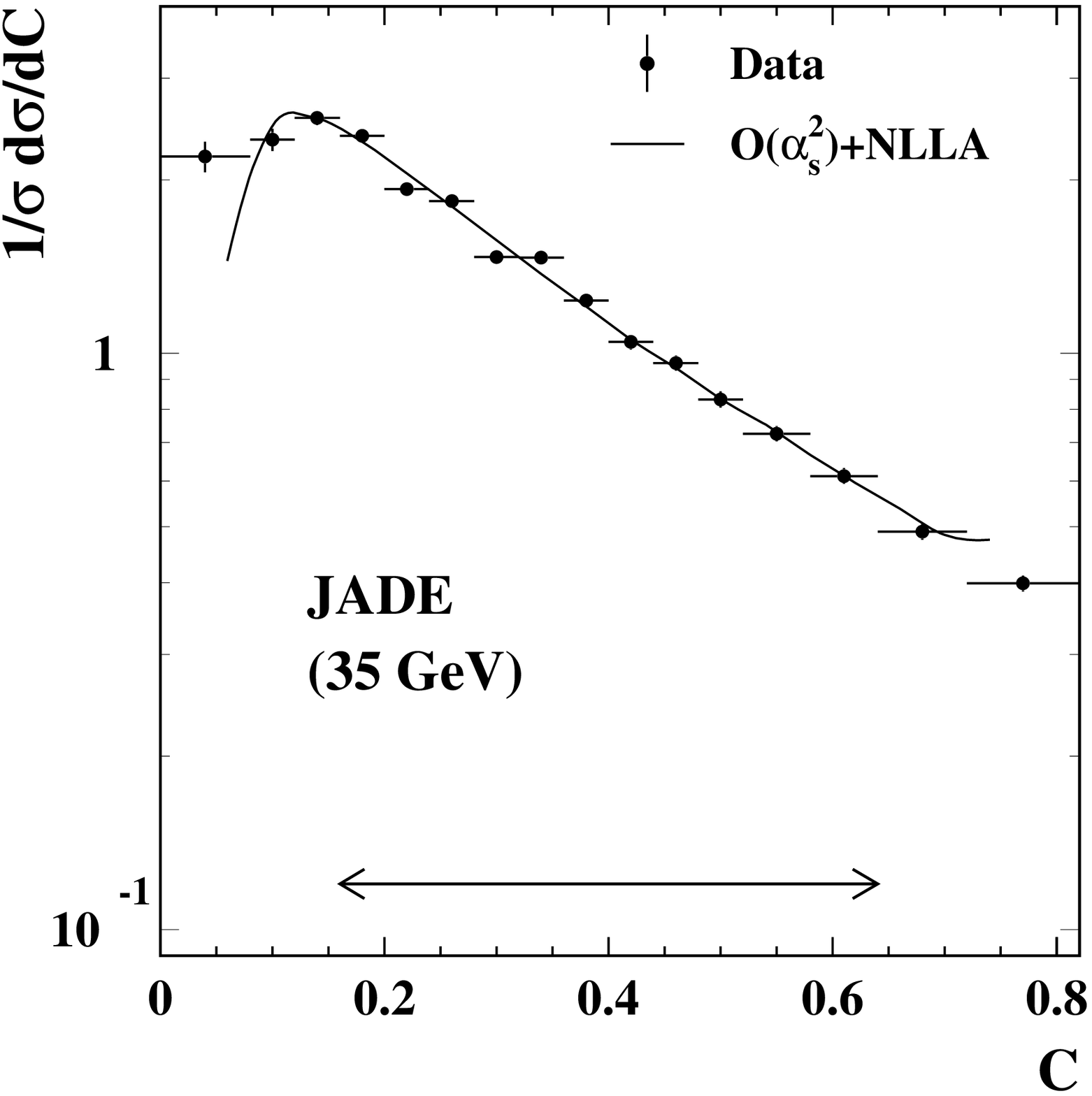}}
\end{center}
\caption{\label{fig-asresult-35GeV}
The same distributions as in Figure~\protect\ref{fig-asresult-44GeV} but for 
$\protect\sqrt{s} = 35$~GeV. 
}
\end{figure}

The fit ranges for each observable were determined   
by choosing the largest range for which the hadronisation
uncertainties remained 
below about $10$~\%, 
for which the \chisqd\ of the fits did not
exceed the minimum by more than a factor of two, 
and by aiming at results for \as\ that are independent from 
changes of the fit range. The remaining changes when      
enlarging or reducing the fit range by one bin on either side   
were taken as systematic uncertainties.
Only statistical errors were considered in the fit thus resulting in   
\chisqd\ larger than unity.                                            
The finally selected fit ranges, the \as\ results of the \chisq\ fits and of 
the study of systematic uncertainties are 
tabulated in Tables~\ref{tab-asresult-44GeV} and \ref{tab-asresult-35GeV}, 
and are shown in Figures~\ref{fig-asresult-44GeV} and \ref{fig-asresult-35GeV}.

We also changed the renormalisation
scale factor in the range of $\xmu = 0.5$ to $2.0$. 
We found variations larger than the 
uncertainties from the detector correction and the 
hadronisation model dependence. 
The dependence of the 
fit result for \as\ on \xmu\ indicates the importance of higher order 
terms in the theory. 

It should be pointed out that the improved perturbative calculation
for the jet broadening resulted in a larger $\alpha_s$ value from the fit to 
the data. The systematic uncertainties are not affected by the new calculation
but the $\chi^2$ improved slightly.

We combined these new results with the \as\ results of our previous
publication~\cite{bib-newJADE} replacing the results obtained from
the jet broadening observables. A single \as\ value was obtained
from the individual determinations from thrust, heavy jet mass, 
total and wide jet broadening, $C$-parameter and differential 2-jet
rate following the procedure described in 
References~\cite{bib-OPALresummed,bib-eventshapes,bib-globalalphas}.
This procedure accounts for correlations of the systematic uncertainties. 
At each energy, a weighted average of the six \as\ values was calculated
with the reciprocal of the square of the respective total error used
as a weight. 
In the case of asymmetric errors we took 
the average 
of the positive and negative error 
to determine the weight.
For each of the systematic checks, the mean of the \as\ values from all
considered observables was determined. 
Any deviation
of this mean from the weighted average of the main result was taken as
a  systematic uncertainty.

With this procedure we obtained as final results for \as
\begin{eqnarray*}
\as(35\ {\mathrm{GeV}}) & = & 0.1448 \pm 0.0010{\mathrm{(stat.)}}
                         \ ^{+0.0117}
                           _{-0.0069}{\mathrm{(syst.)}} \\
\as(44\ {\mathrm{GeV}}) & = & 0.1392 \pm 0.0017{\mathrm{(stat.)}}
                         \ ^{+0.0104}
                           _{-0.0072}{\mathrm{(syst.)}} \ .
\end{eqnarray*}
The systematic errors at $35$ and $44$~GeV are 
the quadratic sums of
the experimental 
uncertainties ($\pm 0.0017$, $\pm 0.0032$), the effects due
to the Monte Carlo modelling 
($^{+0.0070}_{-0.0035}$, $^{+0.0050}_{-0.0027}$) 
and the contributions due to the variation of the renormalisation scale 
($^{+0.0092}_{-0.0057}$, $^{+0.0086}_{-0.0058}$). 
It should be noted that the modelling uncertainties 
due to quark mass effects contribute significantly to the total error.

\section{Mean Values of Distributions and QCD Power Corrections}
\label{sec-meanvalues}
\subsection{ Power corrections} 
The value of \as\ can also be assessed by the energy dependence of 
mean values of event shape distributions. Perturbative calculations
exist for the mean values of thrust, heavy jet mass, total and wide
jet broadening and $C$-parameter up to \oaa. An observable $\cal F$ 
is given by the expression
\begin{displaymath}
   \langle {\cal F}^{\mathrm{pert.}} \rangle = 
                              A_{\cal F}  \left( \frac{\as}{2\pi}\right)  +
              (B_{\cal F} - 2 A_{\cal F}) 
              \left( \frac{\as}{2\pi}\right)^2
\end{displaymath}
where the coefficients $A_{\cal F}$ and $B_{\cal F}$ were determined from the 
\oaa\ perturbative calculations~\cite{bib-ERT,bib-NLLA-1,bib-LEP1report,bib-EVENT2}. 
The term $-2 A_{\cal F}$ accounts for the difference between the total 
cross-section used in the measurement and the Born level cross-section used in 
the perturbative calculation. 
\begin{table}
\begin{center}
\begin{tabular}{|c||c|r||c|
}
\hline
Observable $\cal F$ & $A_{\cal F}$ & \multicolumn{1}{c||}{$B_{\cal F}$} 
                                                  & $a_{\cal F}$  
                                                                         \\
\hline\hline
$\langle T \rangle$       
                    & $2.103$      & $44.99$      & $-1$         
                                                                         \\
$\langle M_H^2/s \rangle$ 
                    & $2.103$      & $23.24$      & $0.5$        
                                                                         \\
$\langle B_T \rangle$ 
                    & $4.066$      & $64.24$      & $0.5$        
                                                                         \\
$\langle B_W \rangle$ 
                    & $4.066$      & $-9.53$      & $0.25$       
                                                                         \\
$\langle C   \rangle$ 
                    & $8.638$      & $146.8$      & $3\pi/2$     
                                                                         \\
\hline
\end{tabular}
\end{center}
\caption{\label{tab-powcor}
Coefficients of the perturbative 
prediction~\protect\cite{bib-ERT,bib-NLLA-1,bib-LEP1report,bib-EVENT2} and coefficients and
parameters of the power corrections~\protect\cite{bib-Milan-factor} to the mean values of 
the event shape observables. Note that the definition of $a_{\cal F}$ does not include
the $2{\cal M}/\pi$ factor introduced in~\protect\cite{bib-Milan-factor}.
}
\end{table}
The numerical values of these coefficients are summarised in Table~\ref{tab-powcor}.

In this study we corrected for hadronisation effects by additive power-suppressed
corrections ($1/\sqrt{s}$) to the perturbative predictions of the 
mean values of the event shape observables.
The non-perturbative effects are due to the emission of very low energetic 
gluons which can not be treated perturbatively due to the divergence of the 
perturbative expressions for \as\ at low scales. 
In the calculations of Reference~\cite{bib-webber} and also \cite{bib-Milan-factor}
which we used in this analysis a non-perturbative
parameter
\begin{displaymath}
\bar{\alpha}_0(\mu_I) = 
   \frac{1}{\mu_I} 
   \int_0^{\mu_I} {\mathrm{d}}k\ \ \as(k)
\end{displaymath}
was introduced to replace the divergent portion of the perturbative
expression for $\as(\sqrt{s})$ below an infrared matching scale $\mu_I$.  
The general form of the power correction to the mean value of an observable 
$\cal F$ was first given in Reference~\cite{bib-webber}. It was the result of
a one-loop calculation. In References~\cite{bib-Milan-factor} similar calculations
have been performed at two-loops. It could be shown 
that the two-loop result modifies the one-loop result only by a factor 
${\cal M} \approx 1.8$
which is known at the Milan factor. An additional factor $2/\pi$ is due to the 
different definitions of the non-perturbative parameters $\alpha_{\mathrm{eff}}$,
being the so-called ``effective coupling''~\cite{bib-effective-coupling}, and
$\bar{\alpha}_0$ as defined above. 

The two-loop result for the power correction assumes for thrust, heavy jet mass
and $C$-parameter the form~\cite{bib-Milan-factor}
\begin{eqnarray*}
   \langle {\cal F}^{\mathrm{pow.}} \rangle & = & 
   a_{\cal F} \frac{4 C_F}{\pi} \cdot \frac{2{\cal M}}{\pi} \cdot
   \left( \frac{\mu_I}{\sqrt{s}}\right) \cdot
                                                  \nonumber \\
    & & 
   \cdot
   \left[
   \bar{\alpha}_{0}(\mu_I)-\as(\sqrt{s}) 
         - \frac{\beta_0}{2\pi}
  \left(\ln\frac{\sqrt{s}}{\mu_I}+\frac{K}{\beta_0}
                                  +1\right)\as^2(\sqrt{s})
   \right]\ ,
\end{eqnarray*}
where $C_F = 4/3$,
while for the two jet broadening measures the correction is logarithmically 
enhanced by a factor $\ln(\sqrt{s}/Q_B)$. 
We approximated $Q_B$ by $\mu_I$ such that
\begin{eqnarray*}
   \langle {\cal F}^{\mathrm{pow.}} \rangle & = & 
   a_{\cal F} \frac{4 C_F}{\pi} \cdot \frac{2{\cal M}}{\pi} \cdot
   \left( \frac{\mu_I}{\sqrt{s}}\right) \cdot
   \ln\left(\frac{\sqrt{s}}{\mu_I}\right) \cdot
                                                  \nonumber \\
    & & 
   \cdot
   \left[
   \bar{\alpha}_{0}(\mu_I)-\as(\sqrt{s}) 
         - \frac{\beta_0}{2\pi}
  \left(\ln\frac{\sqrt{s}}{\mu_I}+\frac{K}{\beta_0}
                                  +1\right)\as^2(\sqrt{s})
   \right]\ .
\end{eqnarray*}
Using this approximation an extra correction without the logarithmic enhancement
\cite{bib-Milan-factor}
was left out
because it is small and strongly anti-correlated to the enhanced correction.
Furthermore we found that the available data on the jet broadening are not 
yet sensitive to this extra correction.

The factor 
$\beta_0 = (11 C_A-2 N_f)/3$ in the two expressions stems from the QCD 
$\beta$-function of the renormalisation group equation. It depends on the 
number of colours, $C_A = 3$, and number of active quark flavours $N_f$, 
for which we used $N_f=5$ throughout the analysis. The term 
$K = (67/18-\pi^2/6) C_A - 5/9 \cdot N_f$ originates from the 
choice of the $\overline{\mathrm{MS}}$ renormalisation scheme.
The remaining coefficient $a_{\cal F}$
is listed in Table~\ref{tab-powcor} for the event shapes considered.

\subsection{ Determination of \as\ using power corrections} 
We determined $\as(\mz)$ by \chisq\ fits of the expression
\begin{displaymath} 
 \langle {\cal F}\rangle = \langle {\cal F}^{\mathrm{pert.}} \rangle + 
                           \langle {\cal F}^{\mathrm{pow.}}  \rangle . 
\end{displaymath} 
to the mean values of the five observables, thrust, heavy jet mass, $C$-parameter,
and total and wide jet broadening,
including the measured mean values obtained by other
experiments at different centre-of-mass 
energies~\cite{bib-L3alphas,bib-OPALNLLA,bib-meanvalues,bib-DELPHI-powcor}.
For the 
central values of \as\ from the fits we chose a renormalisation 
scale factor of $\xmu=1$ and an infrared scale of $\mu_I=2$~GeV.
The \chisqd\ of all fits were between $0.8$ ($\langle M_H^2/s\rangle$)
and $4.4$ ($\langle B_T\rangle$).
We estimated the systematic uncertainties by varying \xmu\ from
$0.5$ to $2$ and $\mu_I$ from $1$ to $3$~GeV. 

\begin{figure}[!htb]
\vspace*{-7mm}
\begin{center}
\resizebox{74mm}{!}{\includegraphics{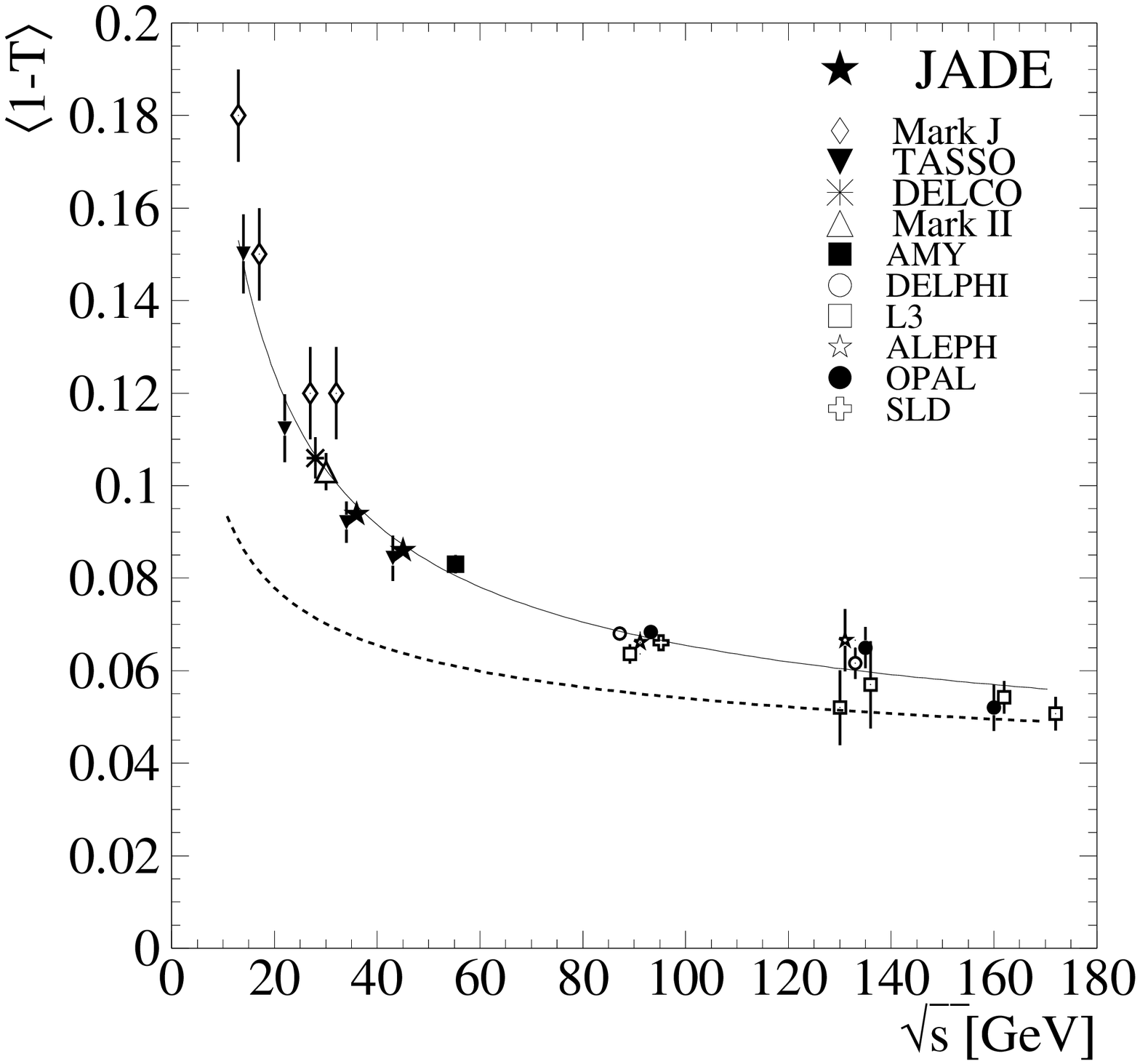}}
\resizebox{74mm}{!}{\includegraphics{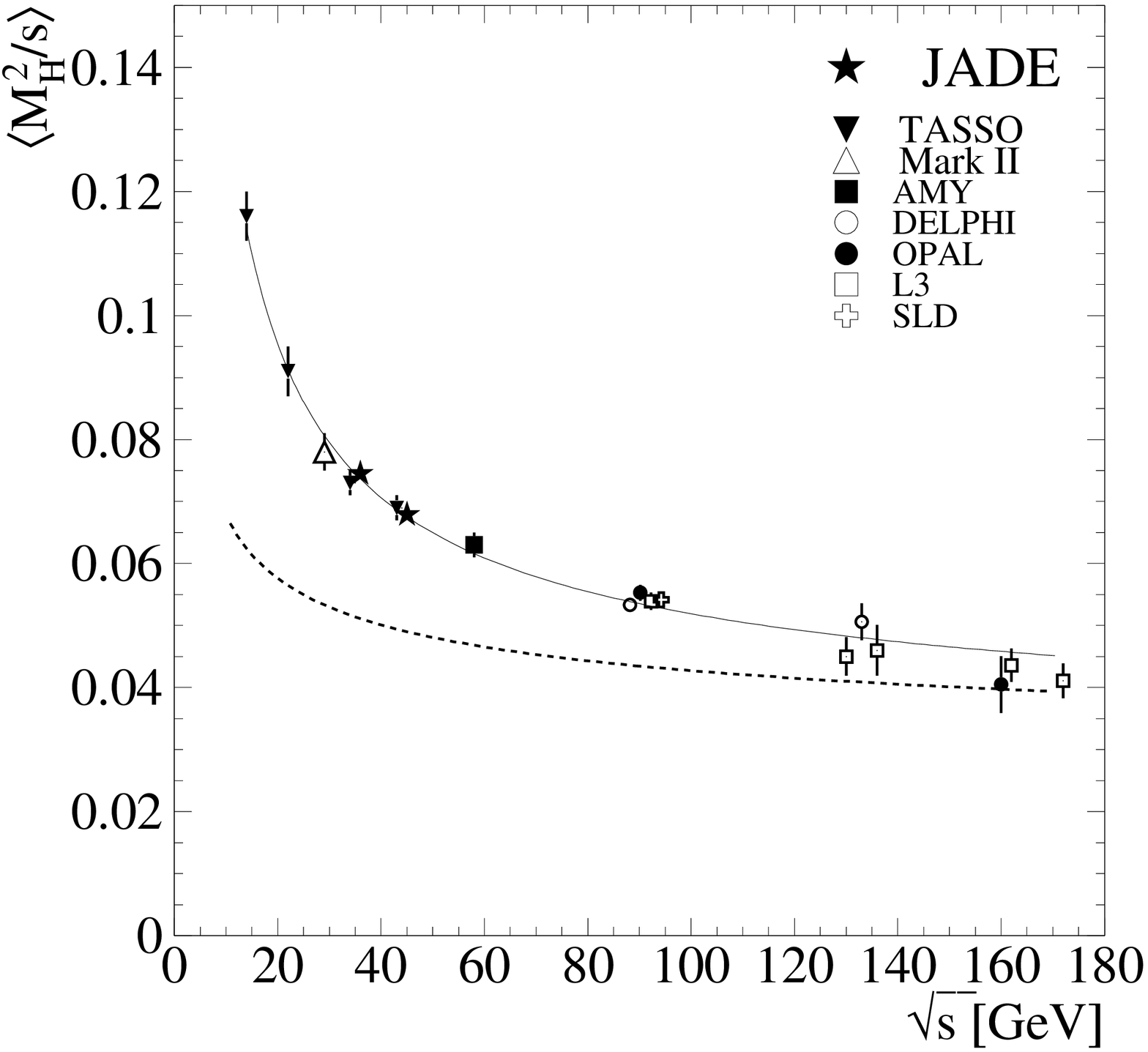}}\\
\resizebox{74mm}{!}{\includegraphics{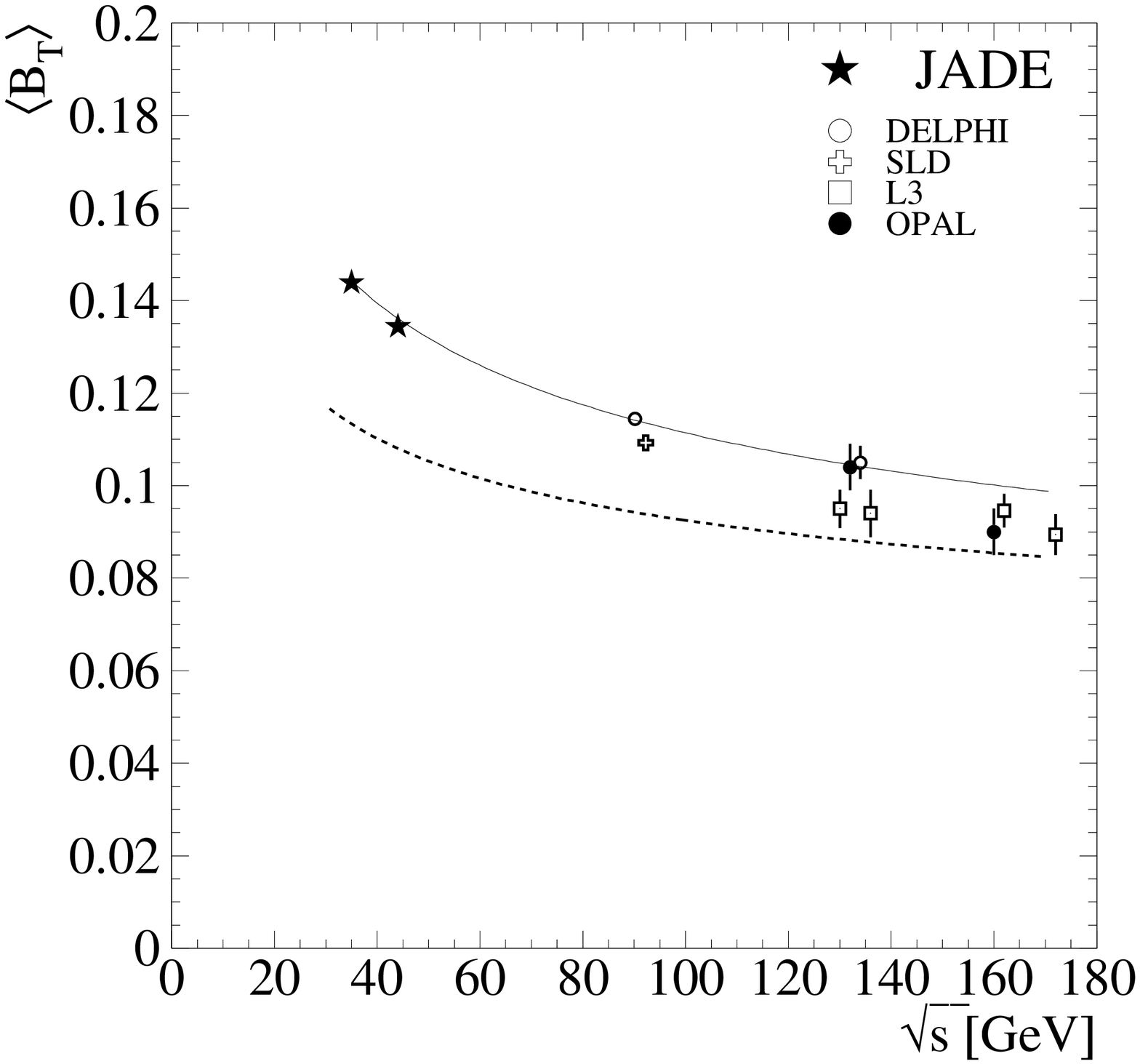}}
\resizebox{74mm}{!}{\includegraphics{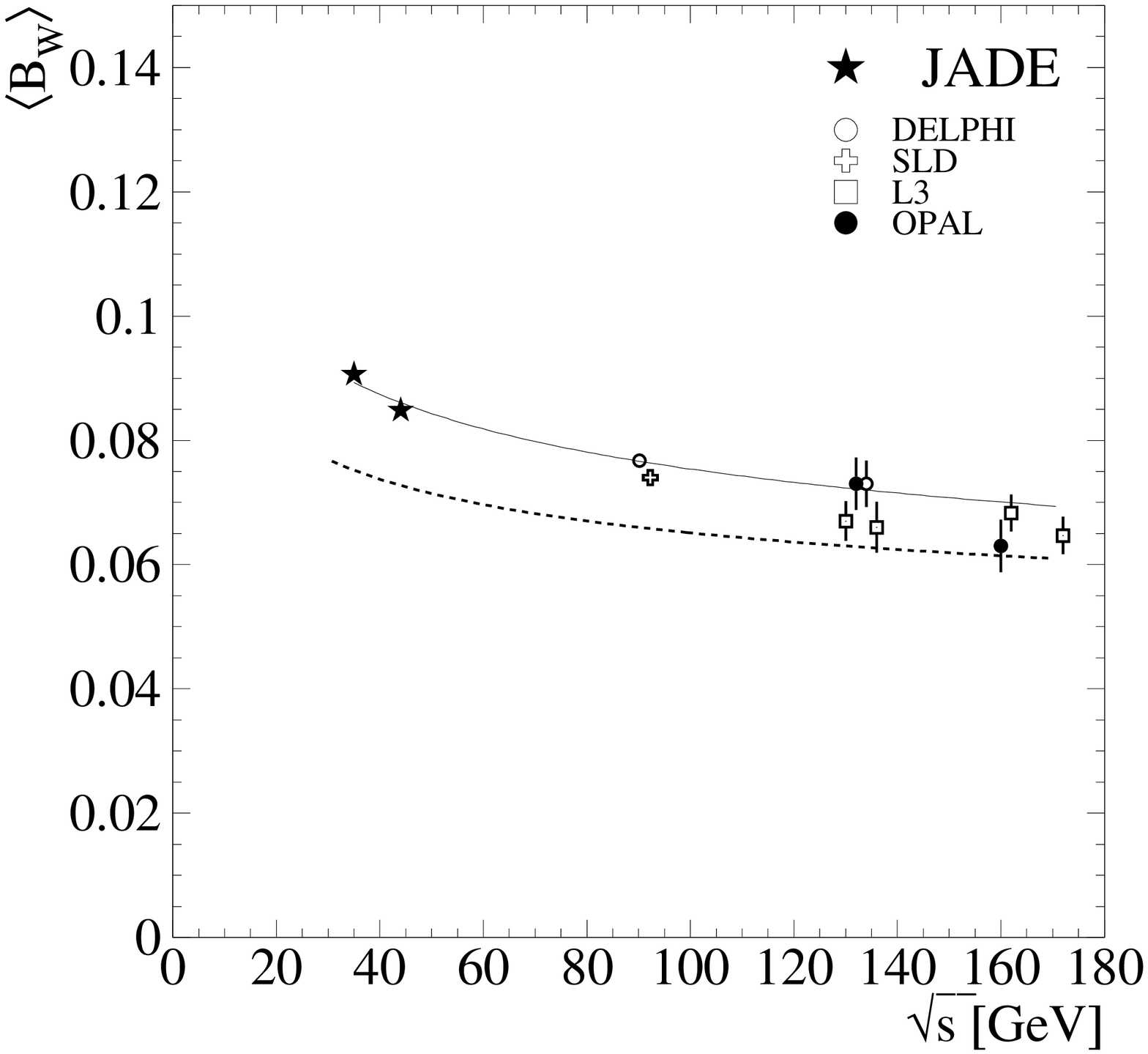}}\\
\resizebox{74mm}{!}{\includegraphics{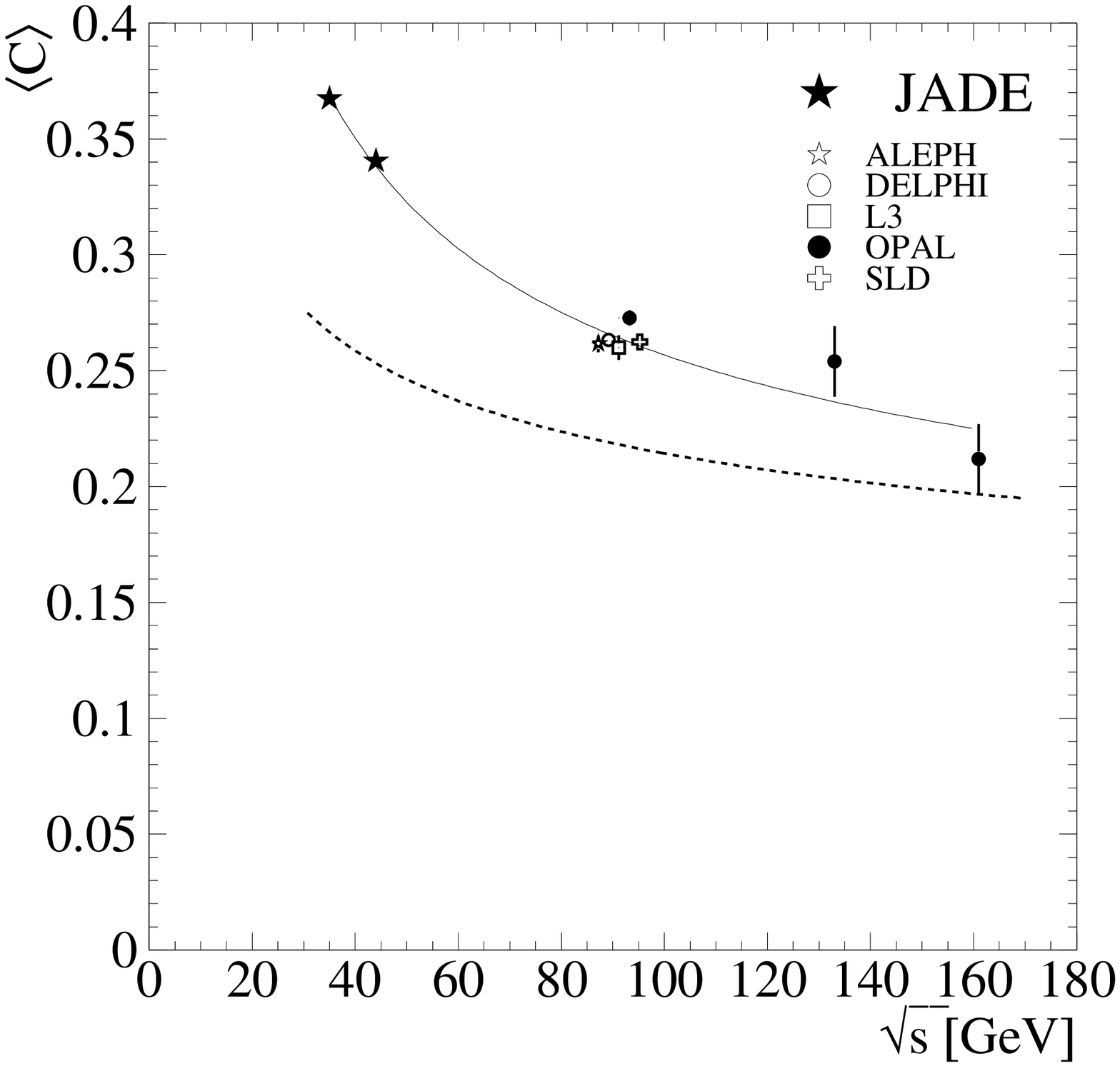}}
\end{center}
\vspace*{-5mm}
\caption{\label{fig-as-powcor}
Energy dependence of the mean values of thrust $\langle 1-T\rangle$,
heavy jet mass $\langle M_H^2/s\rangle$, total $\langle B_T\rangle$
and wide jet broadening $\langle B_W\rangle$, and of the $C$-parameter
$\langle C\rangle$ are 
shown~\protect\cite{bib-L3alphas,bib-OPALNLLA,bib-meanvalues,bib-DELPHI-powcor}. 
The solid curve is the result
of the fit using perturbative calculations plus two-loop power corrections 
which include the Milan factor \protect\cite{bib-Milan-factor} while
the dashed line is the perturbative prediction for the same value of
\asmz.} 
\end{figure}

%
The results of the fits are shown in Figure~\ref{fig-as-powcor}
and the numeric values are tabulated in Table~\ref{tab-as-powcor}.
It presents the values for \as\ and for $\bar{\alpha}_0$, the
experimental errors and systematic uncertainties of the fit
results.
We consider these results based on power corrections as a test of the new
theoretical prediction~\cite{bib-Milan-factor}. 
It should be noted that the theoretically expected 
universality of $\bar{\alpha}_0$ is observed only at the level
of $30\%$. 

Employing the procedure used in Section~\ref{sec-alphas} to combine the 
individual \as\ values, we obtained
\begin{displaymath}
   \as(\mz) = 0.1188\ ^{+0.0044}_{-0.0034}
\end{displaymath}
where the error is the experimental uncertainty ($\pm 0.0016$), the
renormalisation scale uncertainty ($^{+0.0033}_{-0.0023}$) and
the uncertainty due to the choice of the infrared scale 
($^{+0.0024}_{-0.0019}$),
all combined in quadrature. This result is in good agreement with the
world average value~\cite{bib-world-alphas-sb} of 
$\as^{\mathrm{w.a.}}(\mz)=0.119\pm 0.006$.

\section{Summary and Conclusions }
\label{sec-conclusions}
Data recorded by the JADE experiment at centre-of-mass
energies around $35$ and $44$~GeV were analysed in terms
of event shape distributions.

The measured distributions were corrected for detector 
and initial state photon radiation effects using original 
Monte Carlo simulation data 
for $35$ and $44$~GeV. 
The simulated data are based
on the JETSET parton shower generator version~6.3. 
The same event 
generator was also employed to correct the data for hadronisation 
effects in order to determine the strong coupling constant \as.

Our measurements of \as\ are based on the most complete theoretical 
calculations available to date. For all observables 
theoretical calculations exist in \oaa\ %
and in the next-to-leading log approximation.
These two calculations were combined using the ln($R$)-matching scheme.
We found the improved perturbative calculation of the jet broadening to 
describe the data slightly better. With these calculations values of
$\alpha_s$ are obtained which are about $3\%$ higher than previously. 
The $\alpha_s$
values were also more consistent with those from other event shape 
observables.

Combining the values of \as\ obtained in this analysis with those
from our previous publication and using the new values obtained from
jet broadening, the final values 
at the two centre-of-mass energies are
\begin{eqnarray*}
\as(44\ {\mathrm{GeV}}) & = & 0.139 
                         \ ^{+0.011}
                           _{-0.007} \\
\as(35\ {\mathrm{GeV}}) & = & 0.145 
                         \ ^{+0.012}
                           _{-0.007} \ ,
\end{eqnarray*}
where the errors are formed by adding in quadrature the statistical, 
experimental systematics, Monte Carlo modelling and higher order QCD 
uncertainties. The dominant contributions to the total error came from 
the choice of the renormalisation scale and from uncertainties due to 
quark mass effects.

Evolving our \as\ measurements to $\sqrt{s} = \mz$ the results obtained at
$35$ and $44$ ~GeV transform to $0.123\,^{+0.008}_{-0.005}$ and
$0.123\,^{+0.008}_{-0.006}$, respectively. 
The combination of these values gives $\asmz = 0.123\,^{+0.008}_{-0.005}$. 

The energy dependence of the mean values of the distributions can be
directly compared with analytic QCD predictions plus power corrections
for hadronisation effects involving an universal non-perturbative
parameter $\bar{\alpha}_0$~\cite{bib-webber, bib-Milan-factor}.
Our studies resulted in 
\begin{displaymath}
   \as(\mz) = 0.119\ ^{+0.004}
                     _{-0.003}
\end{displaymath}
which is
in good agreement with our results from the \oaa+NLLA fits
and also
with the world average value.
The universality of the non-perturbative parameter $\bar{\alpha}_0$
is found only at a level of $30$\%.

\medskip
\bigskip\bigskip\bigskip
\appendix

%
\clearpage
\section*{ Tables }
\begin{table}[!htb]
%
%
\begin{minipage}[t]{8cm}
\begin{tabular}{|c||r@{ $\pm$ }l@{ $\pm$ }l|}   
\hline
$35 {\mathrm{~GeV}} $           & 
      \multicolumn{3}{c|}{$1/\sigma \cdot {\mathrm{d}}\sigma/{\mathrm{d}}C$} \\
\hline\hline
$ 0.00 $-$0.08  $&$  0.064  $&$   0.004  $&$   0.034 $\\
$ 0.08 $-$0.12  $&$  0.434  $&$   0.019  $&$   0.079 $\\
$ 0.12 $-$0.16  $&$  1.383  $&$   0.039  $&$   0.068 $\\
$ 0.16 $-$0.20  $&$  2.436  $&$   0.056  $&$   0.068 $\\
$ 0.20 $-$0.24  $&$  2.678  $&$   0.060  $&$   0.229 $\\
$ 0.24 $-$0.28  $&$  2.845  $&$   0.062  $&$   0.260 $\\
$ 0.28 $-$0.32  $&$  2.289  $&$   0.054  $&$   0.208 $\\
$ 0.32 $-$0.36  $&$  2.327  $&$   0.056  $&$   0.269 $\\
$ 0.36 $-$0.40  $&$  1.837  $&$   0.047  $&$   0.109 $\\
$ 0.40 $-$0.44  $&$  1.441  $&$   0.041  $&$   0.131 $\\
$ 0.44 $-$0.48  $&$  1.254  $&$   0.038  $&$   0.066 $\\
$ 0.48 $-$0.52  $&$  1.024  $&$   0.034  $&$   0.071 $\\
$ 0.52 $-$0.58  $&$  0.839  $&$   0.025  $&$   0.034 $\\
$ 0.58 $-$0.64  $&$  0.702  $&$   0.022  $&$   0.095 $\\
$ 0.64 $-$0.72  $&$  0.532  $&$   0.017  $&$   0.030 $\\
$ 0.72 $-$0.82  $&$  0.453  $&$   0.014  $&$   0.051 $\\
$ 0.82 $-$1.00  $&$  0.090  $&$   0.004  $&$   0.010 $\\
\hline\hline
mean value       &$  0.3673 $&$     0.0013 $&$    0.0040 $\\
\hline
\end{tabular}
\end{minipage}
\hspace*{0pt\hfill}
\begin{minipage}[t]{8cm}
\begin{tabular}{|c||r@{ $\pm$ }l@{ $\pm$ }l|}   
\hline
$44 {\mathrm{~GeV}}$           & 
      \multicolumn{3}{c|}{$1/\sigma \cdot {\mathrm{d}}\sigma/{\mathrm{d}}C$} \\
\hline\hline
$ 0.00 $-$0.08  $&$   0.093 $&$     0.008 $&$     0.031   $\\
$ 0.08 $-$0.12  $&$   0.801 $&$     0.047 $&$     0.131   $\\
$ 0.12 $-$0.16  $&$   1.953 $&$     0.086 $&$     0.250   $\\
$ 0.16 $-$0.20  $&$   3.016 $&$     0.116 $&$     0.189   $\\
$ 0.20 $-$0.24  $&$   3.172 $&$     0.123 $&$     0.288   $\\
$ 0.24 $-$0.28  $&$   2.759 $&$     0.116 $&$     0.141   $\\
$ 0.28 $-$0.32  $&$   2.457 $&$     0.108 $&$     0.286   $\\
$ 0.32 $-$0.36  $&$   1.748 $&$     0.086 $&$     0.098   $\\
$ 0.36 $-$0.40  $&$   1.623 $&$     0.082 $&$     0.230   $\\
$ 0.40 $-$0.44  $&$   1.163 $&$     0.068 $&$     0.150   $\\
$ 0.44 $-$0.48  $&$   1.159 $&$     0.070 $&$     0.210   $\\
$ 0.48 $-$0.52  $&$   0.847 $&$     0.056 $&$     0.071   $\\
$ 0.52 $-$0.58  $&$   0.720 $&$     0.041 $&$     0.070   $\\
$ 0.58 $-$0.64  $&$   0.626 $&$     0.038 $&$     0.046   $\\
$ 0.64 $-$0.72  $&$   0.503 $&$     0.030 $&$     0.050   $\\
$ 0.72 $-$0.82  $&$   0.348 $&$     0.022 $&$     0.034   $\\
$ 0.82 $-$1.00  $&$   0.079 $&$     0.008 $&$     0.015   $\\
\hline\hline
mean value       &$  0.3404 $&$     0.0023 $&$    0.0037 $\\
\hline
\end{tabular}
\end{minipage}
%
\caption[dummy]{\label{tab-eventshapes-35+44GeV}
Event shape data at $\protect\sqrt{s}=35$ (left) and $44$~GeV (right) 
for the $C$-parameter observable.
The values were corrected for detector 
and for initial state radiation effects. 
The first error denotes the statistical and the second 
the experimental systematic uncertainty.
}
\end{table}

%

\newpage

\begin{table}
\vspace*{-7mm}
\begin{center}
\begin{tabular}{|r||r|r|r|}   \hline
 &\multicolumn{1}{c|}{$B_T$}
 &\multicolumn{1}{c|}{$B_W$} &\multicolumn{1}{c|}{$C$} \\ 
\hline\hline
$\alpha_s$($44$~GeV)
                    &\bf 0.1458 &\bf 0.1318 &\bf 0.1470 \\
\hline\hline
fit range  &$0.08$-$0.27$ &$0.06$-$0.16$ &$0.16$-$0.64$ \\
\hline\hline
$\chi^2$/d.o.f.
                    & $ 3.1 $ & $11.5 $ & $ 1.8 $ \\
\hline\hline
Statistical error
                    & $\pm 0.0014 $ & $\pm 0.0016 $ & $\pm 0.0017 $ \\
\hline\hline
tracks only
                    & $ -0.0027 $ & $ -0.0004 $ & $ -0.0018 $ \\
\hline
clusters only
                    & $ +0.0009 $ & $ +0.0015 $ & $ +0.0015 $ \\
\hline
$\cos\theta_t $
                    & $\pm 0.0005 $ & $\pm 0.0008 $ & $\pm 0.0004 $ \\
\hline
$ p_{\mathrm{miss}} $
                    & $\pm 0.0001 $ & $\pm 0.0002 $ & $\pm 0.0003 $ \\
\hline
$p_{\mathrm{bal}} $
                    & $\pm 0.0002 $ & $\pm 0.0004 $ & $\pm 0.0003 $ \\
\hline
$N_{\mathrm{ch}} $
                    & $ +0.0003 $ & $ +0.0003 $ & $ +0.0003 $ \\
\hline
$E_{\mathrm{vis}} $
                    & $\pm 0.0003 $ & $\pm 0.0002 $ & $\pm 0.0005 $ \\
\hline
fit range
                    & $\pm 0.0023 $ & $\pm 0.0047 $ & $\pm 0.0019 $ \\
\hline\hline
Experimental syst.
                    & $\pm 0.0037 $ & $\pm 0.0051 $ & $\pm 0.0027 $ \\
\hline\hline
$a-0.225$
                    & $ +0.0017 $ & $ +0.0010 $ & $ +0.0026 $ \\
\hline
$a+0.225$
                    & $ -0.0017 $ & $ -0.0009 $ & $ -0.0027 $ \\
\hline
$\sigma_q-30$~MeV
                    & $ +0.0009 $ & $ +0.0007 $ & $ +0.0014 $ \\
\hline
$\sigma_q+30$~MeV
                    & $ -0.0010 $ & $ -0.0006 $ & $ -0.0014 $ \\
\hline
LUND symmetric
                    & $ +0.0017 $ & $ +0.0015 $ & $ +0.0025 $ \\
\hline
$Q_0+500$~MeV
                    & $ -0.0007 $ & $ +0.0012 $ & $ -0.0007 $ \\
\hline
$Q_0-500$~MeV
                    & $ +0.0002 $ & $ -0.0002 $ & $ +0.0002 $ \\
\hline
$\Lambda-50$~MeV
                    & $ -0.0013 $ & $ +0.0001 $ & $ -0.0015 $ \\
\hline
$\Lambda+50$~MeV
                    & $ +0.0008 $ & $ < 0.0001 $ & $ +0.0010 $ \\
\hline
udsc only
                    & $ +0.0064 $ & $ +0.0047 $ & $ +0.0058 $ \\
\hline
MC statistics
                    & $\pm 0.0008 $ & $\pm 0.0009 $ & $\pm 0.0011 $ \\
\hline\hline
\raisebox{2mm}{MC modelling}
                                  &$\stackrel{\textstyle  +0.0072}{-0.0032}$
                                       &$\stackrel{\textstyle  +0.0054}{-0.0026}$
                                            &$\stackrel{\textstyle  +0.0073}{-0.0044}$ \\
\hline\hline
$x_{\mu}=0.5$
                    & $ -0.0100 $ & $ -0.0061 $ & $ -0.0107 $ \\
\hline
$x_{\mu}=2.0$
                    & $ +0.0127 $ & $ +0.0081 $ & $ +0.0133 $ \\
\hline
\hline\hline
\raisebox{2mm}{Total error}
                                  &$\stackrel{\textstyle  +0.0151}{-0.0112}$
                                       &$\stackrel{\textstyle  +0.0111}{-0.0085}$
                                            &$\stackrel{\textstyle  +0.0155}{-0.0120}$ \\
\hline
\end{tabular}
\end{center}
\caption[dummy]{\label{tab-asresult-44GeV}
Values of \as($44$~GeV) derived using the \oaa+NLLA QCD calculations
with $\xmu=1$ and the $\ln(R)$-matching scheme, fit ranges and \chisqd\ %
values for each of the three event shape observables. In addition, the
statistical and systematic uncertainties are given. Where a signed value is
quoted, this indicates the direction in which \as($44$~GeV) changed with
respect to the standard analysis. The scale uncertainty and quark mass
effects are treated as asymmetric uncertainties of \as. 
}
\end{table}

\newpage

\begin{table}[!htb]
\vspace*{-7mm}
\begin{center}
\begin{tabular}{|r||r|r|r|}   \hline
 &\multicolumn{1}{c|}{$B_T$}
 &\multicolumn{1}{c|}{$B_W$} &\multicolumn{1}{c|}{$C$} \\ 
\hline\hline
$\alpha_s$($35$~GeV)
                   &\bf 0.1489 &\bf 0.1367 &\bf 0.1480 \\
\hline\hline
fit range      &$0.08$-$0.27$ &$0.06$-$0.16$ &$0.16$-$0.64$ \\
\hline\hline
$\chi^2$/d.o.f.
                   & $3.1 $ & $4.1 $ & $2.7 $ \\
\hline\hline
Statistical error
                    & $\pm 0.0008 $ & $\pm 0.0009 $ & $\pm 0.0009 $ \\
\hline\hline
tracks only
               & $ -0.0010 $ & $ -0.0006 $ & $ -0.0011 $ \\
\hline
clusters only
                    & $ -0.0009 $ & $ -0.0018 $ & $ -0.0007 $ \\
\hline
$\cos\theta_T $
                    & $\pm 0.0001 $ & $\pm 0.0002 $ & $\pm 0.0002 $ \\
\hline
$ p_{\mathrm{miss}} $
                    & $\pm 0.0001 $ & $\pm 0.0003 $ & $\pm 0.0002 $ \\
\hline
$p_{\mathrm{bal}} $
                    & $\pm 0.0002 $ & $\pm 0.0006 $ & $\pm 0.0002 $ \\
\hline
$N_{\mathrm{ch}} $
                    & $ +0.0005 $ & $ +0.0006 $ & $ +0.0007 $ \\
\hline
$E_{\mathrm{vis}} $
                    & $\pm 0.0001 $ & $\pm 0.0001 $ & $\pm 0.0001 $ \\
\hline
fit range
                    & $\pm 0.0008 $ & $\pm 0.0016 $ & $\pm 0.0004 $ \\
\hline\hline
Experimental syst.
                    & $\pm 0.0014 $ & $\pm 0.0026 $ & $\pm 0.0014 $ \\
\hline\hline
$a-0.225$
                   & $ +0.0023 $ & $ +0.0018 $ & $ +0.0037 $ \\
\hline
$a+0.225$
                    & $ -0.0021 $ & $ -0.0020 $ & $ -0.0030 $ \\
\hline
$\sigma_q-30$~MeV
                    & $ +0.0013 $ & $ +0.0015 $ & $ +0.0017 $ \\
\hline
$\sigma_q+30$~MeV
                    & $ -0.0012 $ & $ -0.0013 $ & $ -0.0015 $ \\
\hline
LUND symmetric
                    & $ +0.0027 $ & $ +0.0029 $ & $ +0.0031 $ \\
\hline
$Q_0+500$~MeV
                    & $ -0.0014 $ & $ +0.0014 $ & $ -0.0004 $ \\
\hline
$Q_0-500$~MeV
                    & $ +0.0006 $ & $ -0.0008 $ & $ +0.0001 $ \\
\hline
$\Lambda-50$~MeV
                    & $ -0.0021 $ & $ -0.0003 $ & $ -0.0024 $ \\
\hline
$\Lambda+50$~MeV
                    & $ +0.0018 $ & $ +0.0003 $ & $ +0.0020 $ \\
\hline
udsc only
                    & $ +0.0086 $ & $ +0.0077 $ & $ +0.0078 $ \\
\hline
MC statistics
                    & $\pm 0.0007 $ & $\pm 0.0008 $ & $\pm 0.0008 $ \\
\hline\hline
\raisebox{2mm}{MC modelling}
                                  &$\stackrel{\textstyle  +0.0099}{-0.0048}$
                                       &$\stackrel{\textstyle  +0.0089}{-0.0045}$
                                            &$\stackrel{\textstyle  +0.0097}{-0.0058}$ \\
\hline\hline
$x_{\mu}=0.5$
                    & $ -0.0107 $ & $ -0.0075 $ & $ -0.0110 $ \\
\hline
$x_{\mu}=2.0$
                    & $ +0.0136 $ & $ +0.0096 $ & $ +0.0138 $ \\
\hline
\hline\hline
\raisebox{2mm}{Total error}
                                  &$\stackrel{\textstyle  +0.0169}{-0.0119}$
                                       &$\stackrel{\textstyle  +0.0134}{-0.0091}$
                                            &$\stackrel{\textstyle  +0.0169}{-0.0126}$ \\
\hline
\end{tabular}
\end{center}
\caption[dummy]{\label{tab-asresult-35GeV}
Values of \as($35$~GeV) derived as in Table~\protect\ref{tab-asresult-44GeV}
but at $35$~GeV.
}
\end{table}

\newpage
\begin{table}[!htb]
\vspace*{-7mm}
\begin{center}
\begin{tabular}{|r||r|r|r|r|r||r|}   \hline
  \multicolumn{1}{|c||}{(a)} 
 &\multicolumn{1}{c|}{$\langle 1-T \rangle$} 
 &\multicolumn{1}{c|}{$\langle M_H^2/s \rangle$}
 &\multicolumn{1}{c|}{$\langle B_T \rangle$}  
 &\multicolumn{1}{c|}{$\langle B_W \rangle$}
 &\multicolumn{1}{c||}{$\langle C \rangle$}
 &\multicolumn{1}{|c|}{average}     \\
\hline\hline
\asmz  
    &\bf 0.1197  &\bf 0.1140  &\bf 0.1211  &\bf 0.1217  &\bf 0.1201  &\bf 0.1188 \\
\hline\hline
$Q$ range [GeV]
    & $13$-$172$ & $14$-$172$ & $35$-$172$ & $35$-$172$ & $35$-$172$ & \\
\hline\hline
\chisqd       
    & $43.2/24$  & $12.1/16$  & $38.9/9$   & $21.6/9$   & $11.0/7$    & \\
\hline\hline
experimental
    &$\pm 0.0013$&$\pm 0.0010$&$\pm 0.0020$&$\pm 0.0020$&$\pm 0.0016$&$\pm 0.0016$ \\
\hline\hline
$\xmu=0.5$    
    &  $-0.0050$ &  $-0.0026$ &  $-0.0039$ &  $-0.0002$ &  $-0.0046$ &  $-0.0023$  \\
\hline
$\xmu=2.0$     
    &  $+0.0061$ &  $+0.0037$ &  $+0.0049$ &  $+0.0011$ &  $+0.0057$ &  $+0.0033$  \\
\hline\hline
$\mu_I=1$~GeV   
    &  $+0.0026$ &  $+0.0013$ &  $+0.0044$ &  $+0.0027$ &  $+0.0025$ &  $+0.0024$  \\
\hline
$\mu_I=3$~GeV 
    &  $-0.0020$ &  $-0.0011$ &  $-0.0036$ &  $-0.0021$ &  $-0.0020$ &  $-0.0019$  \\
\hline\hline
\raisebox{2mm}{Total error}       
    &$\stackrel{\textstyle +0.0068}{-0.0055}$ 
           &$\stackrel{\textstyle +0.0040}{-0.0030}$ 
                  &$\stackrel{\textstyle +0.0069}{-0.0057}$
                         &$\stackrel{\textstyle +0.0035}{-0.0029}$
                                &$\stackrel{\textstyle +0.0064}{-0.0053}$ 
                                       &$\stackrel{\textstyle +0.0044}{-0.0034}$ \\
\hline
\multicolumn{7}{c}{\vspace*{7mm}} \\
  \cline{1-6}
  \multicolumn{1}{|c||}{(b)} 
 &\multicolumn{1}{c|}{$\langle 1-T \rangle$} 
 &\multicolumn{1}{c|}{$\langle M_H^2/s \rangle$}
 &\multicolumn{1}{c|}{$\langle B_T \rangle$}  
 &\multicolumn{1}{c|}{$\langle B_W \rangle$}
 &\multicolumn{1}{c|}{$\langle C   \rangle$} \\
  \cline{1-6} \cline{1-6}
$\bar{\alpha}_0$
    &\bf 0.510   &\bf 0.616   &\bf 0.396   &\bf 0.325   &\bf 0.443   
 &\multicolumn{1}{c} {}                                 \\
  \cline{1-6} \cline{1-6}
experimental
    &$\pm 0.012 $&$\pm 0.018 $&$\pm 0.019 $&$\pm 0.020 $ &$\pm 0.011 $ 
 &\multicolumn{1}{c} {}                                 \\
  \cline{1-6} \cline{1-6}
$\xmu=0.5$    
    &  $+0.004 $ &  $+0.012 $ &  $+0.005 $ &  $+0.082 $ &  $+0.006 $ 
 &\multicolumn{1}{c} {}                                 \\
  \cline{1-6}
$\xmu=2.0$     
    &  $-0.002 $ &  $-0.005 $ &  $-0.001 $ &  $-0.036 $ &  $-0.003 $ 
 &\multicolumn{1}{c} {}                                 \\
  \cline{1-6} \cline{1-6}
\raisebox{2mm}{Total error}       
    &$\stackrel{\textstyle +0.013 }{-0.012 }$
           &$\stackrel{\textstyle +0.022 }{-0.019 }$
                  &$\stackrel{\textstyle +0.020 }{-0.019 }$
                         &$\stackrel{\textstyle +0.084 }{-0.041 }$ 
                                &$\stackrel{\textstyle +0.013 }{-0.011 }$ 
 &\multicolumn{1}{c} {}                                 \\
  \cline{1-6}
\end{tabular}
\end{center}
\caption{\label{tab-as-powcor}
Values of \asmz\ (a) and $\bar{\alpha}_0$ (b) derived 
using $\mu_I=2$~GeV and $\xmu=1$ and 
the \oaa\ calculations and two-loop power corrections 
which include the Milan factor \protect\cite{bib-Milan-factor}.
Fit ranges and 
\chisqd\ values for each of the five event shape observables are specified. 
In addition, the
statistical and systematic uncertainties are given. Where a signed value is
quoted, this indicates the direction in which 
\asmz\ and 
$\bar{\alpha}_0$ changed 
with respect to the standard analysis. The renormalisation and infrared scale 
uncertainties 
are 
treated as an asymmetric uncertainty on 
\asmz. 
These uncertainties are treated 
equally for $\bar{\alpha}_0$ but exclude the infrared scale uncertainty. 
}
\end{table}

%
\clearpage
\section*{ Figures }

\newpage

\begin{thebibliography}{99}

\bibliographystyle{zphysc+}

\bibitem{bib-naroska}
B. Naroska:                         {\em $e^+e^-$ Physics with the JADE
                                    Detector at PETRA},
                                    Phys. Rep. {\bf 148} (1987) 67.

\bibitem{bib-newJADE}
P.~Movilla Fern\'andez, O.~Biebel, S.~Bethke, S.~Kluth, P.~Pfeifenschneider and
the JADE collaboration:            Eur. Phys. J. {\bf C1} (1998) 461.

\bibitem{bib-JADEdet}
JADE Coll., W.~Bartel et~al.:       Phys. Lett. {\bf 88B} (1979) 171.

\bibitem{bib-new-jetbroadening}
Yu.L.~Dokshitzer, A.~Lucenti, G.~Marchesini, G.P.~Salam:
                                 {\em On the QCD analysis of Jet Broadening},
                                 J. High Energy Phys. JHEP {\bf 01} (1998) 011,
                                 IFUM-602-FT,
                                 hep-ph/9801324.

\bibitem{bib-C-resummation}
S.~Catani, B.R.~Webber:         {\em Resummed $C$-Parameter Distribution in
                                     $e^+e^-$-Annihilation},
                                CERN-TH/98-14, Cavendish-HEP-97/16,
                                hep-ph/9801350.

\bibitem{bib-Milan-factor} 
Yu.L.~Dokshitzer, A.~Lucenti, G.~Marchesini, G.P.~Salam,
                                 Nucl. Phys. {\bf B511} (1998) 396;\\
Yu.L.~Dokshitzer, A.~Lucenti, G.~Marchesini, G.P.~Salam:
                                 {\em On the universality of the Milan factor for $1/Q$ 
                                      power corrections to jet shapes},
                                 J. High Energy Phys. JHEP {\bf 05} (1998) 003,
                                 IFUM-601-FT,
                                 hep-ph/9802381;\\
G.P.~Salam:                     {\em The Milan factor for jet-shape observables},
                                 IFUM-623-FT,
                                 hep-ph/9805323.

\bibitem{bib-JADEtrigger}
JADE Coll., W.~Bartel et~al.:       Phys. Lett. {\bf 129B} (1983) 145.

\bibitem{bib-JADEeventsel}
JADE Coll., S.~Bethke et~al.:       Phys. Lett. {\bf B213} (1988) 235.

%
%
%
%
\bibitem{bib-OPALresummed}
OPAL Coll., P.D.~Acton et~al.:      Z. Phys. {\bf C59} (1993) 1.

%
%
%
%

%
\bibitem{bib-JETSET}
T.~Sj\"ostrand:                     Comput. Phys. Commun. {\bf 39} (1986) 347;\\
T.~Sj\"ostrand, M.~Bengtsson:       Comput. Phys. Commun. {\bf 43} (1987) 367.

\bibitem{bib-C-parameter}
G.~Parisi,                     Phys. Lett. {\bf 74B} (1978) 65;\\
J.F.~Donoghue, F.E.~Low, S.Y.~Pi, Phys. Rev. {\bf D20} (1979) 2759.

\bibitem{bib-JADEtune}
JADE Coll., E. Elsen et~al.:        Z. Phys. {\bf C46} (1990) 349.

%
%
%
%
\bibitem{bib-OPALtune}
OPAL Coll., G.~Alexander et~al.:    Z. Phys. {\bf C69} (1995) 543.

\bibitem{bib-Peterson}
C.~Peterson, D.~Schlatter, I.~Schmitt, P.M.~Zerwas:
                                    Phys. Rev. {D27} (1983) 105.

\bibitem{bib-eventshapes}
OPAL Coll., G.~Alexander et~al.:    Z. Phys. {\bf C72} (1996) 191.

\bibitem{bib-globalalphas}
OPAL Coll., P.D.~Acton et~al.:       Z. Phys. {\bf C55} (1992) 1.

\bibitem{bib-ERT} 
R.K.~Ellis, D.A.~Ross, A.E.~Terrano: Nucl. Phys. {\bf B178} (1981) 421.

\bibitem{bib-NLLA-1}
S.~Catani, G.~Turnock, B.R.~Webber:  Phys. Lett. {\bf B295} (1992) 269.

\bibitem{bib-LEP1report}
Z.~Kunszt, P.~Nason, G.~Marchesini, B.R.~Webber:
                                    in {\em Z Physics at LEP 1}, vol. 1,
                                    ed. G.~Altarelli, R.~Kleiss and C.~Verzegnassi,
                                    CERN Yellow Book 89-08.

\bibitem{bib-EVENT2}
We used the program EVENT2 of S.~Catani and M.H.~Seymour to 
determine the perturbative coefficients by integrating the ERT \oaa\ 
matrix elements~\cite{bib-ERT}:\\
S.~Catani, M.H.~Seymour:            Phys. Lett. {\bf B378} (1996) 287.

%
\bibitem{bib-webber}
Yu.L.~Dokshitzer, B.R.~Webber:      Phys. Lett. {\bf B352} (1995) 451;\\
B.R.~Webber:                        proceedings of the
                                    {\em Workshop on Deep Inelastic
                                         Scattering and QCD (DIS 95)},
                                    Paris, France, 24-28 Apr, 1995,
                                    ed. J.F.~Laporte and Y.~Sirois;
                                    Cavendish-HEP-95/11; hep-ph/9510283.

\bibitem{bib-effective-coupling}
Yu.L.~Dokshitzer, G.~Marchesini, B.R.~Webber: Nucl. Phys. {\bf B469} (1996) 93.

\bibitem{bib-L3alphas}
L3 Coll., M.~Acciarri et~al.:        Phys. Lett. {\bf B404} (1997) 390.

\bibitem{bib-OPALNLLA} 
OPAL Coll., K.~Ackerstaff et~al.:   Z. Phys. {\bf C75} (1997) 193.

\bibitem{bib-meanvalues}
ALEPH Coll., D.~Buskulic et~al.:     Z. Phys. {\bf C55} (1992) 209;\\
ALEPH Coll., D.~Buskulic et~al.:     Z. Phys. {\bf C73} (1997) 409;\\
AMY   Coll., Y.K.~Li et~al.:         Phys. Rev. {\bf D41} (1990) 2675;\\
DELCO Coll., M.~Sakuda et~al.:       Phys. Rev. {\bf 152B} (1985) 399;\\
DELPHI Coll., P.~Abreu et~al.:       Z. Phys. {\bf C73} (1996) 11;\\
L3 Coll., B.~Adeva et~al.:           Z. Phys. {\bf C55} (1992) 39;\\
L3 Coll., M.~Acciarri et~al.:        Phys. Lett. {\bf B371} (1996) 137;\\
MarkII Coll., A.~Peterson et~al.:    Phys. Rev. {\bf D37} (1988) 1;\\
MarkJ Coll., D.P.~Barber et~al.:     Phys. Rev. Lett. {\bf 43} (1979) 901;\\
MarkJ Coll., D.P.~Barber et~al.:     Phys. Lett. {\bf 85B} (1979) 463;\\
OPAL Coll., P.D.~Acton et~al.:       Z. Phys. {\bf C55} (1992) 1;\\
OPAL Coll., G.~Alexander et~al.:     Z. Phys. {\bf C72} (1996) 191;\\
SLD Coll., K.~Abe et~al.:            Phys. Rev. {\bf D51} (1995) 962;\\
TASSO Coll., W.~Braunschweig et~al.: Z. Phys. {\bf C41} (1988) 359;\\
TASSO Coll., W.~Braunschweig et~al.: Z. Phys. {\bf C47} (1990) 187.

\bibitem{bib-DELPHI-powcor}
DELPHI Coll., P.~Abreu et~al.:       Z. Phys. {\bf C73} (1997) 229.

\bibitem{bib-world-alphas-sb}
S.~Bethke:                          proceedings of the 
                                    {\em QCD Euroconference 97},
                                     Montpellier, France, July 3-9
                                    (1997), Nucl. Phys. B (Proc.Suppl.) 64
                                     (1998) 54-62;
                                     hep-ex/9710030.


%
%
%
%
%
%

\end{thebibliography}
\end{document}